\documentstyle[aps,twocolumn]{revtex}
\begin{document}
\draft
\title{Gravitational energy as Noether charge}
\author{Sean A. Hayward}
\address{Center for Gravitational Physics and Geometry,
104 Davey Laboratory, The Pennsylvania State University,
University Park, PA 16802-6300, U.S.A.\\
{\tt hayward@gravity.phys.psu.edu}}
\date{13th April 2000}
\maketitle

\begin{abstract}
A definition of gravitational energy is proposed 
for any theory described by a diffeomorphism-invariant Lagrangian.
The mathematical structure is a Noether-current construction of Wald 
involving the boundary term in the action,
but here it is argued that 
the physical interpretation of current conservation is conservation of energy.
This leads to a quasi-local energy defined for compact spatial surfaces.
The energy also depends on a vector generating a flow of time.
Angular momentum may be similarly defined, 
depending on a choice of axial vector.
For Einstein gravity: 
for the usual vector generating asymptotic time translations,
the energy is the Bondi energy;
for a stationary Killing vector, the energy is the Komar energy;
in spherical symmetry, for the Kodama vector, 
the energy is the Misner-Sharp energy.
In general, 
the lack of a preferred time indicates the lack of a preferred energy,
reminiscent of the energy-time duality of quantum theory.
\end{abstract}

\section{Introduction}

The notion of gravitational energy is perhaps 
the most outstanding unresolved conceptual issue in Einstein gravity.
Energy is of fundamental importance in most branches of physics,
yet there is no accepted definition of the energy of the gravitational field.
Indeed, the general view is that 
the equivalence principle forbids a local gravitational energy density.
Nevertheless, the total energy of an asymptotically flat space-time 
as defined by the Bondi energy is accepted as physically meaningful, 
since it satisfies a certain conservation law 
involving the energy flux of gravitational waves, e.g.\cite{W1}.
Similarly, there are definitions of energy in spherical\cite{sph,1st} 
and cylindrical symmetry\cite{cyl} 
which have the desired physical properties,
again including conservation laws.
All these definitions are quasi-local in the sense that 
they depend on a spatial surface 
(with respect to the conformal metric in the asymptotic case) 
rather than a spatial hypersurface; 
in other words, they are surface rather than volume integrals.
There have been many attempts to construct such quasi-local energies,
but as yet no consensus.

Energy in classical or quantum physics 
is most commonly understood via an action.
For instance, in Einstein gravity, 
the energy-momentum-stress tensor of the matter may be defined by 
the variation of the matter Lagrangian with respect to the metric.
The energy-momentum-stress tensor of the gravitational field 
may similarly be defined by the variation of the Einstein-Hilbert Lagrangian 
with respect to the metric, leading to the Einstein tensor, 
with the Einstein equation expressing zero combined energy.
This makes sense physically, but this local energy does not lead to 
the type of quasi-local energy discussed above;
it gives a definition of the mass of the matter as a volume integral,
with no contribution from gravitational energy.
Any variation of any Lagrangian will similarly give zero combined energy 
by the corresponding field equation.

However, this involves the usual assumption that 
boundary terms in the variation are ignored.
One might expect that 
these boundary terms give a boundary contribution to energy, 
to be added to the vanishing bulk terms.
This is the idea explored in this article.
Similar ideas have been proposed by other authors, 
e.g.\ recently\cite{E,FFFR,CN} and references therein,
but with different implementation and results.
In particular, this author regards as very basic criteria that
the energy should recover the Schwarzschild mass 
for the Schwarzschild black hole 
and be generally well defined and real inside black holes.
The desired physical properties of gravitational energy 
are perhaps best illustrated by the symmetric cases\cite{sph,1st,cyl}.

A general framework for studying boundary terms 
in diffeomorphism-invariant actions has been developed 
by Wald\cite{W2} and Iyer \& Wald\cite{IW} in a different context,
namely to try to define black-hole entropy.
The key idea is that one can always construct a Noether current 
which is conserved when the field equations hold.
Here it will be argued that the physical meaning of this is 
conservation of energy, when the diffeomorphism generates a flow of time.
The corresponding Noether charge then furnishes a quasi-local energy.

\section{Boundary terms in the action}

The method may be summarized as follows,
taking the case of a 4-manifold $M$ for definiteness.
The dynamical fields are a set of tensor fields on $M$ 
which will be denoted simply by $\phi$.
When there is a metric $g$, the remaining fields will be denoted by $\psi$.
Then the particular dynamical theory is defined by a Lagrangian 4-form $L[\phi]$
which is invariant under diffeomorphisms:
\begin{equation}
L[\xi^*(\phi)]=\xi^*L[\phi]
\end{equation}
where $\xi^*$ denotes 
the action of the diffeomorphism generated by a vector $\xi$.
In particular, Einstein gravity is defined by
\begin{equation}
L[g,\psi]={{*}R[g]\over{16\pi}}+L_m[g,\psi]
\end{equation}
where $R$ is the Ricci scalar and $*$ the Hodge dual,
mapping $p$-forms to $(4-p)$-forms,
with ${*}1$ the space-time volume 4-form.
The first term is the Einstein-Hilbert Lagrangian,
with units such that the Newton constant is unity,
and the second term is the matter Lagrangian, 
with $\psi$ being the matter fields.
Then the variations
\begin{equation}
{\delta\over{\delta g}}({*}R)\cong-{*}G\qquad
{\delta L_m\over{\delta g}}\cong{{*}T\over2}
\end{equation}
define the Einstein tensor $G$ 
and the energy-momentum-stress tensor $T$ in contravariant form,
where $\cong$ denotes equality up to boundary terms.
Then the action principle
\begin{equation}
\delta\int_M L=0
\end{equation}
yields the Einstein equation
\begin{equation}
G=8\pi T.
\end{equation}
Returning to the general case, the boundary term may be included as
\begin{equation}
\delta L={*}(\Phi\circ\delta\phi)+d\Theta
\end{equation}
where $d$ is the exterior derivative
and $\circ$ denotes the sum of the contractions of each $\phi$ 
with its appropriate ${*}\Phi\cong\delta L/\delta\phi$, 
so that $\Phi$ is a set of tensors dual to $\phi$.
Here the boundary 3-form $\Theta[\phi,\delta\phi]$ 
is called the symplectic potential\cite{IW}.
Then
\begin{equation}
\delta\int_ML=\int_M{*}(\Phi\circ\delta\phi)+\int_{\partial M}\Theta
\end{equation}
and the action principle 
for variations fixed on the boundary $\partial M$ yields the field equations
\begin{equation}
\Phi=0.
\label{field}
\end{equation}
The Noether current defined by Wald\cite{W2,IW} is the 3-form
\begin{equation}
J[\phi,\xi]=\Theta[\phi,{\cal L}_\xi\phi]-\xi\cdot L[\phi]
\label{current}
\end{equation}
where ${\cal L}_\xi$ denotes the Lie derivative along $\xi$
and the dot denotes contraction between last and first indices respectively.
Then the general identity
\begin{equation}
{\cal L}_\xi\Lambda=\xi\cdot d\Lambda+d(\xi\cdot\Lambda)
\label{identity}
\end{equation}
for a $p$-form $\Lambda$ yields
\begin{equation}
dJ=-{*}(\Phi\circ{\cal L}_\xi\phi).
\end{equation}
Thus there is a conservation law
\begin{equation}
dJ=0
\label{law}
\end{equation}
when the field equations (\ref{field}) hold.
Therefore locally there exists a 2-form $Q$ such that
\begin{equation}
J=dQ
\end{equation}
when the field equations hold.
It should be noted that there is gauge dependence in $Q$,
given by
\begin{equation}
L\mapsto L+d\alpha\qquad
\Theta\mapsto\Theta+d\beta\qquad
Q\mapsto Q+d\gamma.
\end{equation}
However, the transformation of $Q$ will not affect the energy,
to be defined as an integral of $Q$ over a compact surface (without boundary).
Also, a standard choice of $L$ is usually given for a particular theory.
For $\Theta$, in practice, natural gauge choices often exist.

\section{Energy as Noether charge}

It will now be proposed that the physical interpretation 
of the conservation law (\ref{law}) is {\em conservation of energy}, 
if $\xi$ generates a flow of time.
Then the 1-form ${*}2J$ is the {\em energy-momentum density} of the boundary.
This can be seen from
\begin{equation}
\delta L={{*}(T:\delta g)\over2}-{{*}(G:\delta g)\over{16\pi}}
+{*}(\Psi\circ\delta\psi)+d\Theta
\end{equation}
where the colon denotes double contraction.
Just as the first term yields the bulk energy-momentum-stress $T$,
so the last term yields the boundary energy-momentum.
The transformation (\ref{current}) from $\Theta$ to $J$ may be regarded as 
a covariant analog of the construction of a Hamiltonian from a Lagrangian.
The unfortunate factor of 2 follows from the conventions of Wald\cite{W2,IW}.
Then $2Q$ is the surface energy density of the boundary.
Taking a spatial surface $S\subset\partial M$, assumed connected and compact, 
the {\em energy} is defined as the Noether charge
\begin{equation}
E=\oint_S2Q.
\end{equation}
It seems natural to take $\xi$ to generate $\partial M$ locally from $S$.
When the field equations hold, 
the energy may also be written as a volume integral
\begin{equation}
E=\int_\Sigma 2J
\end{equation}
where $\Sigma$ is a spatial hypersurface such that $S=\partial\Sigma$.
However, conservation of $J$ means that 
$E$ is independent of the choice of $\Sigma$;  it is really a surface integral.
As such, it is a quasi-local energy, 
though it also depends on the diffeomorphism vector $\xi$, 
which represents a choice of time.
Generally one may denote the dependence as $E_S[\phi,\xi]$, 
though this will be shortened henceforth to $E[\xi]$.

\section{Einstein gravity}

Iyer \& Wald\cite{IW} found that for pure Einstein gravity, $L={*}R/16\pi$,
the gauge freedom may be naturally fixed so that
\begin{equation}
Q[\xi]=-{{*}d\xi\over{8\pi}}.
\end{equation}
The sign depends on the convention for $*$, which is chosen as
\begin{equation}
{*}1=\sqrt{\det g}\,dt\wedge dr\wedge d\vartheta\wedge d\varphi
\end{equation}
where $\xi=\partial/\partial t$ 
and $(\vartheta,\varphi)$ are polar coordinates on $S$,
assumed henceforth of spherical topology.
Then
\begin{equation}
E[\xi]=-{1\over{4\pi}}\oint_S{*}d\xi.
\end{equation}
This may be recognized as the Komar energy $E_{st}$
if $\xi$ is a stationary Killing vector $\xi_{st}$\cite{W1}:
\begin{equation}
E_{st}=E[\xi_{st}].
\end{equation}
It is also a standard expression for the Bondi energy $E_\infty$
if $\xi$ is the vector $\xi_\infty$ generating asymptotic time translations,
gauge-fixed as described e.g.\ by Wald\cite{W1}:
\begin{equation}
E_\infty=E[\xi_\infty].
\end{equation}
A third example is spherical symmetry:
in terms of the areal radius $r$,
there is a preferred flow of time given by the Kodama vector\cite{sph,1st}
\begin{equation}
\xi_{sph}=\hat{*}dr
\end{equation}
where $\hat{*}$ is the Hodge operator of the space normal to the spheres,
covariant and contravariant duals with respect to $g$ 
not being denoted explicitly.
A widely accepted definition of energy is the Misner-Sharp energy\cite{sph,1st}
\begin{equation}
E_{sph}=(1-dr\cdot dr)r/2.
\end{equation}
A recent definition of dynamic surface gravity is\cite{1st}
\begin{equation}
\kappa=\hat{*}d\hat{*}dr/2.
\end{equation}
Then
\begin{equation}
E[\xi_{sph}]=r^2\kappa.
\end{equation}
This can be interpreted as 
a relativistic version of the Newtonian law of gravitation.
One component of the Einstein equation reads\cite{1st}
\begin{equation}
E_{sph}=r^2\kappa-2\pi r^3\hbox{tr}\,T
\label{Einstein}
\end{equation}
where the trace is in the normal space.
Thus in vacuo, 
\begin{equation}
E_{sph}=E[\xi_{sph}].
\end{equation}
This also holds for any matter field for which $\hbox{tr}\,T$ vanishes, 
such as a massless Klein-Gordon field.
Otherwise, one needs to include the contribution from the matter Lagrangian,
as will be seen below for the Einstein-Maxwell case.
The fourth example of cylindrical symmetry\cite{cyl} 
does not work in this simple way,
indicating that 
the definition needs to be modified (or abandoned) for non-spherical topology.
With this caveat,
the known energies are recovered as special cases of the general definition 
when a preferred flow of time exists.

It may also be proposed that $E$ can represent energy in more general senses,
such as momentum and angular momentum, depending on the nature of $\xi$.
For instance, a definition of {\em angular momentum} would be
\begin{equation}
-E[\xi_{ax}]/2
\end{equation}
where $\xi_{ax}$ is an axial vector, particularly an axial Killing vector.
For the Kerr black hole in the usual Boyer-Lindquist coordinates,
one does indeed find
\begin{equation}
m=E[\xi_{st}]\qquad ma=-E[\xi_{ax}]/2
\end{equation}
where $m$ and $ma$ are the usual mass and angular momentum 
respectively\cite{W1}.
This illustrates that there is no need for $\xi$ to be hypersurface-orthogonal.

\section{Klein-Gordon and Maxwell fields}

For the purposes of this article, the variation $\delta$ of the Wald method 
need only ever be a Lie derivative ${\cal L}_\xi$.
Consequently one may write $\Theta$ as a function of $(\phi,\xi)$ 
without explicitly determining it as a function of $(\phi,\delta\phi)$, 
which saves some calculation.
It should be noted that the most natural fixing of the gauge dependence 
may be different by these two methods.
The identity (\ref{identity}) yields
\begin{equation}
{\cal L}_\xi d\Lambda=d{\cal L}_\xi\Lambda.
\end{equation}

Warming up with the Klein-Gordon field $\psi$, defined by the Lagrangian
\begin{equation}
L[g,\psi]=-{*}(d\psi\cdot d\psi+m^2\psi^2)/2
\end{equation}
one finds
\begin{equation}
\delta L={*}(T:\delta g)/2+{*}\Psi d\psi+d\Theta
\end{equation}
where
\begin{eqnarray}
&&T=d\psi\otimes d\psi-(d\psi\cdot d\psi+m^2\psi^2)g/2\\
&&\Psi={*}d{*}d\psi-m^2\psi\\
&&\Theta=-\delta(\psi{*}d\psi)/2.
\end{eqnarray}
The three terms respectively give the energy-momentum-stress tensor,
the Klein-Gordon equation and the energy-momentum density
\begin{equation}
2J=2dQ-\psi\xi\cdot({*}\Psi)
\end{equation}
where
\begin{equation}
2Q=-\psi\xi\cdot({*}d\psi).
\end{equation}
Turning to the Maxwell field, defined by the Lagrangian
\begin{equation}
L[g,A]=-{{*}(F:F)\over{16\pi}}
\end{equation}
where the 1-form $A$ is the electromagnetic potential 
and $F=2dA$ is the electromagnetic field tensor, one finds
\begin{equation}
\delta L={*}(T:\delta g)/2+{*}(\Psi\cdot dA)+d\Theta
\end{equation}
where
\begin{eqnarray}
&&T=-(F\cdot F+(F:F)g/4)/4\pi\\
&&\Psi={*}d{*}F/4\pi\\
&&\Theta=-\delta({*}(A\cdot F))/8\pi.
\end{eqnarray}
The three terms respectively give the energy-momentum-stress tensor,
the Maxwell equation (paired with $dF=0$) and the energy-momentum density
\begin{equation}
2J=2dQ-\xi\cdot({*}(\Psi\cdot A))
\end{equation}
where
\begin{equation}
2Q=-\xi\cdot({*}(A\cdot F))/4\pi.
\end{equation}
This is clearly gauge-dependent, 
but for the Reissner-Nordstr\"om black hole 
there is a natural gauge choice\cite{W1} $A=-(e/r)dt$
which leads to the Maxwell energy
\begin{equation}
E_M=\oint_S2Q_M={e^2\over r}
\end{equation}
where the suffix distinguishes the Maxwell terms from the pure Einstein terms,
given as above by
\begin{equation}
E_E=\oint_S2Q_E=r^2\kappa
\end{equation}
where $\xi=\partial/\partial t$.
Using the component (\ref{Einstein}) of the Einstein equation 
with the appropriate $\hbox{tr}\,T=-e^2/4\pi r^4$ and $E_{sph}=m-e^2/2r$,
the combined Einstein-Maxwell charge is found to be\cite{AH}
\begin{equation}
E_E+E_M=m.
\end{equation}
Remarkably, this is just the usual mass.
Note that this holds anywhere in the space-time, not just on the horizon.
This illustrates that, 
in order to obtain the expected mass in the presence of matter fields,
the appropriate energy is the combined energy 
of the matter and gravitational field, not that of the latter alone.
A similar situation occurs in the Hamiltonian method 
to determine the mass of an isolated horizon\cite{ABF}.

\section{Conclusion}

For any theory defined by a diffeomorphism-invariant Lagrangian, 
an energy has been defined as the Noether charge 
associated with a conserved Noether current, an energy-momentum density
derived from the boundary term in the variation of the action.
The conservation law is interpreted as conservation of energy.
The energy is quasi-local in the sense that 
it is an integral over a compact spatial surface.
It also depends on a vector generating a flow of time through the surface.
This neatly answers the equivalence-principle objection to gravitational energy:
generally there is no preferred time, 
different choices giving different energies.
However, when there is a preferred flow of time, 
such as in a stationary or spherically symmetric space-time, 
or at infinity in an asymptotically flat space-time, 
the definition does indeed recover standard definitions of energy 
for Einstein gravity, due to Komar, Misner \& Sharp and Bondi respectively.
In each case, $E_{known}=E[\hbox{preferred vector}]$.
In the presence of matter, the matter Lagrangian should also be included,
as the Reissner-Nordstr\"om case shows.

Comparing with quantum theory, where energy becomes an operator
\begin{equation}
\hat E=i\hbar\xi^*
\end{equation}
explicitly determined by the chosen flow of time,
the duality of energy and time is a fundamental feature,
expressed for instance by the Heisenberg uncertainty principle.
This article suggests, for a wide class of gravitational theories, 
a purely classical relationship between gravitational energy and time.
Understanding the connection may provide new paths in quantum gravity.

\bigskip\noindent
The author is grateful to Abhay Ashtekar 
and the Center for Gravitational Physics and Geometry for hospitality.
Research supported by the National Science Foundation under award PHY-9800973.


\begin{references}
\bibitem{W1}R M Wald, General Relativity (University of Chicago Press 1984).
\bibitem{sph}S A Hayward, {Phys. Rev.} {\bf D53}, 1938 (1996).
\bibitem{1st}S A Hayward, {Class. Quantum Grav.} {\bf 15}, 3147 (1998).
\bibitem{cyl}S A Hayward, {Class. Quantum Grav.} {\bf 17}, 1749 (2000).
\bibitem{E}R J Epp, 
Angular momentum and an invariant quasilocal energy in general relativity
(gr-qc/0003035).
\bibitem{FFFR}L Fatibene, M Ferraris, M Francaviglia \& M Raiteri,
Noether charges, Brown-York quasilocal energy and related topics
(gr-qc/0003019).
\bibitem{CN}C-N Chen \& J M Nester,
A symplectic Hamiltonian derivation of quasilocal energy-momentum for GR,
(gr-qc/0001088).
\bibitem{W2}R M Wald, {Phys. Rev.} {\bf D48}, 3427 (1993).
\bibitem{IW}V Iyer \& R M Wald, {Phys. Rev.} {\bf D50}, 846 (1994).
\bibitem{AH}M C Ashworth \& S A Hayward, 
Noether currents of charged spherical black holes (in preparation).
\bibitem{ABF} A Ashtekar, C Beetle \& S Fairhurst, 
{Class. Quantum Grav.} {\bf 17}, 253 (2000).
\end{references}
\end{document}